\documentclass[showpacs,floatfix,superscriptaddress]{revtex4}
\usepackage{graphicx}
\usepackage{epsfig}
\usepackage{bm}
\usepackage{placeins}
\usepackage[utf8]{inputenc}

\usepackage[utf8]{inputenc}
\usepackage{amssymb}
\usepackage{float}
\usepackage{amsmath}
\usepackage{dcolumn}
\usepackage{latexsym}
\usepackage{subfig}
\usepackage{amsthm}
\usepackage{array}
\usepackage{framed}
\usepackage{cancel}
\usepackage[colorlinks]{hyperref}
\usepackage[usenames,dvipsnames]{color}
\hypersetup{
     breaklinks=true,
    pdfstartview={FitH},    
    colorlinks=true,       
    linkcolor=blue,          
    citecolor=red,        
    filecolor=magenta,      
    urlcolor=blue,           
    anchorcolor=green,      
    linktocpage=true
}

\def\doi{http://doi.org}

\usepackage{float}

\begin{document}

\title{Shadow of Extreme Compact Charged Objects in Consistent 4-Dimensional Einstein-Gauss-Bonnet Gravity}

\author{Sara Saghafi}
\email[]{s.saghafi@umz.ac.ir }
\author{Kourosh Nozari}
\email[]{ knozari@umz.ac.ir (Corresponding Author)} 
\author{Maryam Kaveh}
\email[]{m.kaveh4697@gmail.com}

\affiliation{Department of Theoretical Physics, Faculty of Science, University of Mazandaran,\\
P. O. Box 47416-95447, Babolsar, Iran}

\begin{abstract}
In order to better describe gravitational phenomena on both very small and cosmological scales, there have been constant attempts to generalize and expand the theory of General Relativity (GR) since its inception. The Einstein–Gauss–Bonnet (EGB) theory is one such extension that adds spacetime corrections related to curvature. Since the standard Gauss–Bonnet term is purely topological, it does not contribute to the field equations in four dimensions. To get around this restriction, however, an invariant four-dimensional limit $(D\to 4)$ has been developed. In this work, we study Extreme Compact Charged Objects (ECCOs), which can resemble black holes, in a gravity framework that is compatible with Einstein-Gauss-Bonnet in four dimensions. Our main goal is to compare theoretical predictions with Event Horizon Telescope (EHT) observational data in order to constrain the Gauss–Bonnet coupling constant $\alpha$. In order to achieve this, we investigate important optical characteristics like the shadow, light-bending angle, and other associated observables, as well as the geodesic structure of ECCO spacetimes in EGB gravity. Finally, we apply these findings to constrain the Gauss-Bonnet constant. 
\vspace{12 pt}
\end{abstract}


\maketitle

\enlargethispage{\baselineskip}
\tableofcontents

\section{Introduction}\label{intro}
Existing observational evidence must continue to be compatible with any coherent alternative theory of gravity. This mainly refers to conformity to the extremely accurate tests in the solar system in the weak-field regime ~\cite{Will:2005va,Will:2018bme}. On the other hand, current observations still allow for significant flexibility in the strong-field domain, which includes the areas around compact astrophysical objects like neutron stars and black holes  ~\cite{Berti:2015itd,Barack:2018yly,CANTATA:2021ktz}. Important insights into the strong-field regime of gravity have been gained from the LIGO/VIRGO collaboration's detection of gravitational waves from black hole and neutron star mergers ~\cite{Abbott_2016,LIGOScientific:2017vwq}. Recent horizon-scale images of the supermassive black holes in the Milky Way and M87 taken by the Event Horizon Telescope have also provided complementary information ~\cite{2019ApJ...875L...1E,EventHorizonTelescope:2022wkp}. However, a number of alternative models of gravity have already been subjected to strict limitations due to high-precision pulsar measurements (see, e.g.,~\cite{Shao:2017gwu,Freire:2022wcz} ). When taken as a whole, these observational discoveries have generated a notable renewed interest in the theoretical investigation of gravitational physics.\\

Black hole formation is predicted by the most successful theory of gravity, general relativity. Additionally, it has sparked interest in investigating other unusual compact objects like naked singularities, boson stars, quark stars and wormholes. Nonetheless, it is extremely unlikely that some of these objects will ever be realized physically within the framework of General Relativity due to their violation of energy conditions. A sizable amount of research has explored the potential existence of such exotic compact objects (see, for example,~\cite{Cardoso:2016oxy,Mark:2017dnq,Ou:2021efv,Cunha:2022gde} ), which encourages further research into compact objects in alternative and modified theories of gravity.\\

Among the higher derivative gravitational theories, the Einstein–Gauss–Bonnet (EGB) gravity is widely recognized to include higher curvature corrections to the Einstein-Hilbert action. The Gauss-Bonnet (GB) term, a topological quantity in four dimensions, does not generically contribute to field equations unless it is accompanied by extra fields. However, a new 4D EGB theory of gravity has been proposed by Glavan and Lin~\cite{Glavan_2020} that avoids the Ostrogradsky instability and the consequences of Lovelock's theorem. Their method was to take the limit $D \rightarrow 4$  after rescaling the Gauss–Bonnet coupling as $\alpha \to \alpha/(D-4)$ . Through this process, they were able to obtain a nontrivial black hole solution, which is now known as the novel four-dimensional EGB theory. The proposal immediately attracted a lot of attention and was extensively investigated in a variety of contexts, such as black holes coupled to magnetic charge or nonlinear electrodynamics \cite{Jusufi:2020qyw,Abdujabbarov:2020jla}, electrically charged black holes \cite{Fernandes_2020_2,Zhang:2020sjh}, and static and spherically symmetric black hole configurations and their related physical properties \cite{Ghosh:2020syx,Konoplya:2020ptx,Singh:2021nvm,Wei:2020poh,Yang:2020jno}. Strong and weak gravitational lensing by black holes \cite{Kumar:2020sag,Jin:2020emq}, quasi–normal mode spectra \cite{Heydari-Fard:2020sib,Mishra:2020gce,Aragon:2020qdc}, black hole shadows \cite{Konoplya:2020bxa,Guo:2020zmf,Zeng:2020dco}, wormholes and thin–shell wormholes \cite{Jusufi:2020yus,Zhang:2020kxz}, and several other related topics \cite{Samart:2020sxj} were the subjects of additional studies in this framework. More recently, the Newman–Janis algorithm has been used to construct rotating extensions of the theory \cite{Kumar:2020owy,NaveenaKumara:2020kpz}.\\

Examining the shadow images of compact objects is very interesting from a theoretical point of view. Such investigations are relevant not only for black holes within General Relativity \cite{Bambi:2012tg,Johannsen:2013vgc,Cunha:2015yba,Cunha:2016bpi,Gralla:2019xty,Younsi:2021dxe,Promsiri:2023rez} and in various modified gravity theories \cite{Zeng:2020dco,Aktar:2024akk,Peng:2020wun,Guo:2020zmf,Kumar:2020owy,Konoplya:2020bxa,Gyulchev:2021dvt,Zeng:2022fdm,Ye:2023qks}, but also for more exotic candidates such as wormholes \cite{Bambi:2013nla,Schee:2021pdt,Bambi:2021qfo,Guerrero:2021pxt,Rahaman:2021web,Tsukamoto:2021fpp,Peng:2021osd,Guerrero:2022qkh,Delijski:2022jjj,Huang:2023yqd,Ishkaeva:2023xny}, naked singularities \cite{Tavlayan:2023vbv,Vagnozzi:2022moj,Gyulchev:2020cvo,Joshi:2020tlq,Dey:2020bgo,Gyulchev:2019tvk,Shaikh:2018lcc,Deliyski:2023gik}, and boson stars \cite{Rosa:2022toh,Rosa:2022tfv}. Recent studies continue to explore these phenomena, including the shadows of black holes like Compact Object in modified theories \cite{Aktar:2024akk}, how parameter constraining can influence the mass accretion process of a black hole in modified theories of gravity \cite{Mukherjee:2024kzq}, and the gravitational lensing by wormholes and naked singularities \cite{Maurya:2025cgn,Ditta:2025wwx} . One important result of previous studies is that some exotic compact objects could replicate black hole-like shadow patterns as their light ring structures are similar. Furthermore, black hole shadows are essential for connecting theory and observation. The shadow has developed into a potent observational tool to examine the near-horizon geometry thanks to the groundbreaking images captured by the Event Horizon Telescope. In addition to providing consistency tests for General Relativity, its exact size and shape enable one to evaluate potential signs of new physics beyond Einstein's theory and to constrain the parameter space of alternative gravity models ~\cite{Vagnozzi:2022moj,Gong:2025sjv,Fathi:2025bvi,Wang:2025vsx,Aliyan:2024xwl,Nozari:2023flq,Kurmanov:2025bvp,Kala:2025fld,Wang:2025buh,Olmo:2025ctf,Cai:2025pan,Nozari:2024jiz} . Additionally, by identifying the differences between classical black holes and other compact objects, shadow analysis can provide information about the characteristics of ultra-compact configurations, their stability, and possible astrophysical applications. As a result, one of the most promising approaches to testing basic physics with strong gravitational fields is now thought to be shadow analyses.\\

In Ref.~\cite{Gammon:2024gij} the authors have obtained a novel solution describing an extremely compact charged object within the framework of four-dimensional Einstein–Gauss–Bonnet (EGB) gravity. They explored several physical aspects of this configuration, including its horizon structure, stability properties, and possible astrophysical relevance. Motivated by these results above, it becomes natural to extend the analysis toward the optical appearance of such objects. In particular, investigating the shadow cast by this compact configuration provides a direct way to connect the theoretical solution with astrophysical observations. Since the Event Horizon Telescope (EHT) has already delivered high-resolution images of the supermassive black hole M87*, the shadow size and shape extracted from these observations serve as a valuable benchmark. Therefore, in the present work we aim to study the shadow characteristics of the charged EGB compact object and confront them with the EHT measurements by performing a detailed comparison between the predicted shadow radius and the observed size of M87*. Such a study not only sheds light on the observational signatures of higher-curvature corrections in gravity but also offers a pathway to constrain the parameter space of EGB gravity through current and future black hole imaging data.\\

This paper is organized as follows: In the second section \eqref{STVG}, the modified four-dimensional Einstein-Gauss-Bonnet gravity theory is introduced  briefly, and then the line element of the dark compact object in the theory is introduced. In the third section \eqref{Bshadow}, we investigate the effective potential, shadow behavior, energy emission rate, and deflection angle of the dark compact object in the four-dimensional Einstein-Gauss-Bonnet gravity. In the fourth section \eqref{ARMECCO}, we establish constraints on the EGB parameter using the Event Horizon Telescope data.  Finally, in the fifth section \eqref{SaC}, we summarize, conclude, and discuss our main results.

\section{Theoretical Framework for Extreme Compact Charged Objects in
 Regularized 4D Einstein-Gauss-Bonnet Gravity}\label{STVG}

The action of four--dimensional Einstein--Gauss--Bonnet (4DEGB) gravity coupled to matter and electromagnetism is of the form,
\begin{equation}
S = \frac{1}{16\pi G}\int d^4x \sqrt{-g}\,\Big(R + \alpha \mathcal{L}_{GB} \Big) + S_{\text{fluid}} + S_{\text{Maxwell}} ,
\end{equation}
where $\alpha$ is the Gauss--Bonnet coupling and $\mathcal{L}_{GB}$ denotes the regularized scalar--Gauss--Bonnet interaction. Variation of this action with respect to the scalar field and the metric yields the modified field equations, which reduce to the Einstein equations in the limit $\alpha \to 0$.  

To describe stellar configurations, in Ref.\cite{Gammon:2024gij}, the authors impose static spherical symmetry and adopt the metric ansatz
\begin{equation}
ds^2 = - e^{\Phi(r)} dt^2 + e^{\Lambda(r)} dr^2 + r^2 d\Omega^2 ,
\end{equation}
with two radial functions $\Phi(r)$ and $\Lambda(r)$. Outside the matter distribution, asymptotic flatness requires $\Phi(r) = -\Lambda(r)$. From the $tt$--component of the field equations one obtains a generalized expression for the radial metric function,
\begin{equation}\label{rrr}
e^{-\Lambda(r)} = 1 + \frac{r^2}{2\alpha}\left[1 - \sqrt{1 + 4\alpha\left(\frac{2m(r)}{r^3} - \frac{q(r)^2}{r^4}\right)} \,\right],
\end{equation}
where the function $m(r)$ naturally arises and is interpreted as the enclosed mass within radius $r$. In the absence of matter, $m(r)$ reduces to the constant ADM mass $M$.  

The inclusion of charge proceeds through the Maxwell action with a conserved current. This introduces the charge function $q(r)$, defined such that
\begin{equation}
E(r) = \frac{q(r)}{r^2},
\end{equation}
is the electric field generated by the enclosed charge. The stress--energy tensor then consists of quark matter described as a perfect fluid together with the electromagnetic field contribution.  

Combining the modified gravitational field equations with Maxwell’s equations yields the generalized Tolman--Oppenheimer--Volkoff (TOV) system in 4DEGB gravity:
\begin{align}
\frac{dq}{dr} &= 4\pi r^2 \rho_e\, e^{\Lambda/2}, \\[6pt]
\frac{dm}{dr} &= 4\pi r^2 \rho(r) + \frac{q(r)}{r}\frac{dq}{dr}, \\[6pt]
\frac{dP}{dr} &= -(\rho(r)+P(r))\,\frac{r^3\!\left(\Gamma + 8\pi\alpha P(r)-1\right) - 2\alpha m(r)}{\Gamma r^2 \left[(\Gamma - 1) r^2 - 2\alpha \right]} 
+ \frac{q(r)}{4\pi r^4}\frac{dq}{dr},
\end{align}
with
\begin{equation}
\Gamma = \sqrt{1 + 4\alpha\left(\frac{2m(r)}{r^3} - \frac{q(r)^2}{r^4}\right)} .
\end{equation}

Here, $\rho(r)$ and $P(r)$ denote the energy density and pressure of quark matter, while $\rho_e$ is the charge density. These equations govern the stellar interior: the first describes the accumulation of charge, the second determines the enclosed mass, and the third is the hydrostatic equilibrium condition modified by Gauss--Bonnet and electromagnetic effects.  

At the stellar surface $r=R$, the pressure vanishes, $P(R)=0$, and the total mass and charge are defined as $M = m(R)$ and $Q = q(R)$. In the vacuum exterior, the metric reduces to the charged 4DEGB black hole solution, ensuring consistency with asymptotic flatness.

\section{Optical Features of ECCO in 4D EGB gravity}\label{Bshadow}

Black holes can deflect light from their path because of their very strong gravity. Some of these rays escape the black hole and some are trapped. These photons that are trapped by the black hole create a dark region in space, which is called the black hole shadow. In this section, we present formulas for the deflection angle, energy emission rate, and shadow shape for a test particle  and  by considering an arbitrary values of $\alpha = \{0.0001,\ 0.125,\ 0.250,\ 0.375,\ 0.500\}$, we study  optical appearance  of the Extreme Compact Charged Object  in four-dimensional Einstein-Gauss-Bonnet gravity. In Ref.~\cite{Gammon:2024gij}, the authors, in a study of charged quark stars and highly compressed bodies in four-dimensional Einstein-Gauss-Bonnet gravity, determined that the Gauss-Bonnet coupling constant lies within the range $0 < \alpha \lesssim 3.2$, which we use  here to choose arbitrary values for $\alpha$.

\subsection{ Null geodesics}

In this part of the paper, our main goal is to investigate the behavior of the  effective potential for a Extreme Compact Charged Object in four-dimensional Einstein-Gauss-Bonnet gravity. For this purpose, we first  introduce the Lagrangian of a test particle in this spacetime as follows:
\begin{equation}
L = \frac{1}{2} g_{ab} \dot{x}^a \dot{x}^b,
\label{eqlagrangian}
\end{equation}
where a  dot denotes differentiation with respect to the affine parameter \(\tau\). The canonically conjugate momentum components corresponding to Eq.~\eqref{eqlagrangian} are:
\begin{equation}
P_t = f(r) \dot{t} = E  \ ,
\label{eq:Pt}
\end{equation}
\begin{equation}
P_r = \frac{1}{f(r)} \dot{r} \  ,
\label{eq:Pr}
\end{equation}
\begin{equation}
P_\theta = r^2 \dot{\theta}  \ ,
\label{eq:Ptheta}
\end{equation}
\begin{equation}
P_\phi = r^2 \sin^2 \theta \dot{\phi} = L  \ ,
\label{eq:Pphi}
\end{equation}
where \(L\) and  \(E\) denote the conserved angular momentum and energy of the test particle as constants of motion associated with the spacetime symmetries, respectively.

We apply the Hamilton-Jacobi method to analyze photon orbits around the Extreme Compact Charged Object. In four-dimensional Einstein-Gauss-Bonnet gravity, the Hamilton-Jacobi method is presented as follows: 
\begin{equation}
\frac{\partial S}{\partial \tau} = -\frac{1}{2} g^{ab} \frac{\partial S}{\partial x^a} \frac{\partial S}{\partial x^b}\ .
\label{eq:Hamilton-Jacobi}
\end{equation}

Substituting the metric components from Eq.~\eqref{rrr} into Eq.~\eqref{eq:Hamilton-Jacobi} results in:
\begin{equation}
-2 \frac{\partial S}{\partial \tau} = -\frac{1}{f(r)} \left( \frac{\partial S}{\partial t} \right)^2 + f(r) \left( \frac{\partial S}{\partial r} \right)^2 + \frac{1}{r^2} \left( \frac{\partial S}{\partial \theta} \right)^2 + \frac{1}{r^2 \sin^2 \theta} \left( \frac{\partial S}{\partial \phi} \right)^2.
\label{eq:hamilton-jacobi-expanded}
\end{equation}

Assuming a separable solution for the action \( S \), we write:
\begin{equation}
S = \frac{1}{2} m^2 \tau - E t + L \phi + S_r(r) + S_\theta(\theta)   \ ,
\label{eq:separable-action}
\end{equation}
where \(m\) is the rest mass of the test particle. For photons, we set \(m=0\). Inserting Eq.~\eqref{eq:separable-action} into Eq.~\eqref{eq:hamilton-jacobi-expanded} yields :
\begin{equation}
0 = \frac{E^2}{f(r)} - f(r) \left( \frac{d S_r}{d r} \right)^2 - \frac{1}{r^2} \left( \frac{d S_\theta}{d \theta} \right)^2 - \frac{L^2}{r^2 \sin^2 \theta}\ ,
\end{equation}

\begin{equation}
\frac{E^{2}}{f(r)} = f(r) \left( \frac{d S_r}{d r} \right)^{2} + \frac{K}{r^{2}}\ .
\end{equation}

Where $K$ is the Carter constant.  Using Eqs.~\eqref{eq:Pt}--\eqref{eq:Pphi}, the equations of motion for the photon (null geodesics) are:
\begin{align}
\dot{t} &= \frac{E}{f(r)}\   , \label{eq:tdot} \\
r^2 \dot{r} &= \pm \sqrt{R}\   , \label{eq:rdot} \\
r^2 \dot{\theta} &= \pm \sqrt{\Theta}\   , \label{eq:thetadot} \\
\dot{\phi} &= \frac{L}{r^2 \sin^2 \theta}  \ , \label{eq:phidot}
\end{align}

where the signs \(+\) and \(-\) denote outgoing and ingoing radial motion, respectively. The quantities  R and  \(\Theta\)  are defined  as  follows:
\begin{align}
R &= r^4 E^2 - r^2 (L^2 + K) f(r) \ , \label{eq:R}\\
\Theta &= K - \frac{L^2}{\sin^2 \theta}\  .\label{eq:Theta}
\end{align}

The path of the photon is given by Eqs.~\eqref{eq:tdot}--\eqref{eq:phidot}. The equation of radial motion for a particle moving in a gravitational field is as follows:
\begin{equation}
\left(\frac{dr}{d\tau}\right)^2 + V_{\mathrm{eff}} = 0 \ ,
\end{equation}
where the effective potential is given by:
\begin{equation}
V_{\mathrm{eff}} = \frac{f(r)}{r^2} (K + L^2) - E^2  \ .
\end{equation}
In the equatorial plane, i.e. when $\theta = \pi/2$, the Carter constant ($K$) reduces to $L^{2}$.
The boundary of the shadow is associated with the unstable circular photon's orbits and is determined by the maximum of the effective potential, given by the photon orbit radius at the radius \(r_0\):
\begin{equation}
V_{\mathrm{eff}} \big|_{r=r_0} = 0  \ , \quad \frac{d V_{\mathrm{eff}}}{d r} \bigg|_{r=r_0} = 0 \ , \quad R \big|_{r=r_0} = 0  \ , \quad \frac{d R}{d r} \bigg|_{r=r_0} = 0  \ . \label{eq:mar}
\end{equation}

Among the possible positive roots of Eq.~\eqref{eq:mar2}, the smallest one corresponds to the radius of the unstable circular photon orbit, denoted by \( r_0 \),  which determines the boundary of the black hole shadow.This radius satisfies the following condition:
\begin{equation}
r_0 f'(r_0) - 2 f(r_0) = 0\label{eq:mar2}\  ,
\end{equation}
where the prime indicates differentiation with respect to the radial coordinate.

Substitution of Eq.\eqref{rrr} into Eq.\eqref{eq:mar} results  in the effective potential  for the charged Extreme Compact Charged Objects in four-dimensional Einstein-Gauss-Bonnet gravity as follows:
\begin{equation}
V_{\mathrm{eff}} = -E^2 + \frac{K + L^2}{r^2}  \left[ 1 + \frac{r^2}{2\alpha} \left( 1 - \sqrt{1 + 4 \alpha \left( -\frac{q^2}{r^4} + \frac{2M}{r^3} \right) } \right) \right] \ .
\label{eq:Veff}
\end{equation}

\begin{figure}[H]
\centering
\subfloat[\label{Fig1a} for q=0.25]{\includegraphics[width=0.475\textwidth]{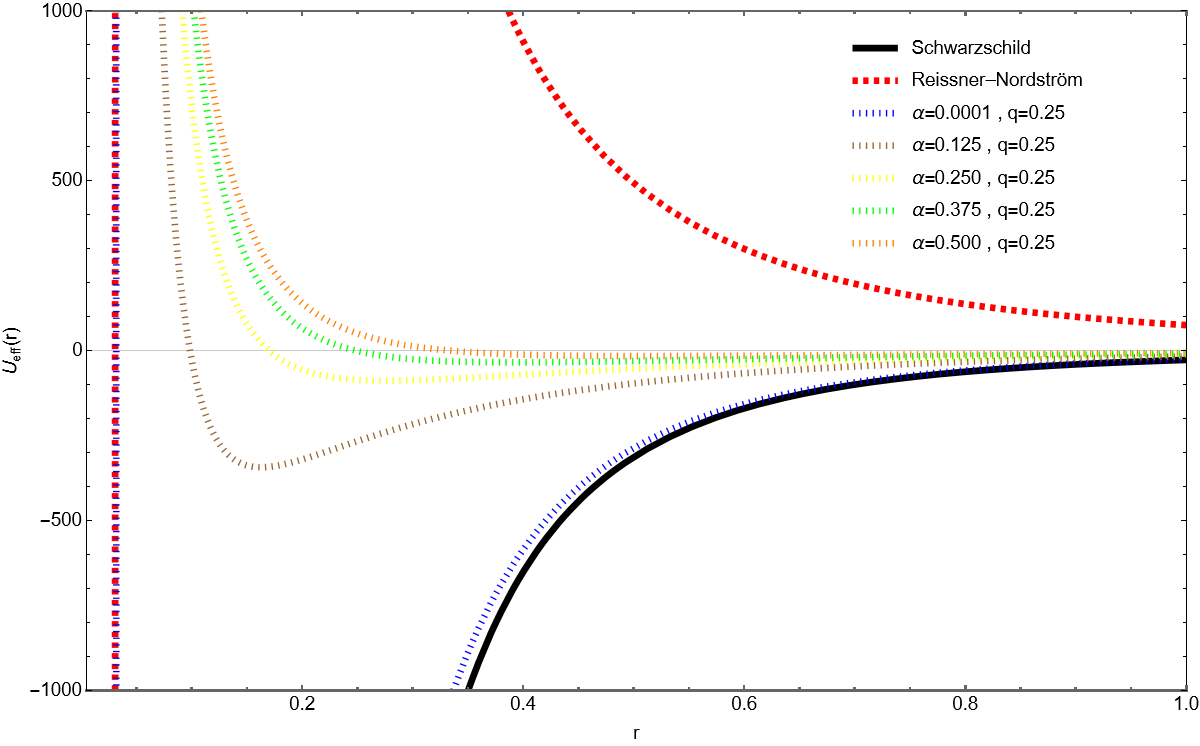}}
\,\,\,
\subfloat[\label{Fig1b} for q=0.5]{\includegraphics[width=0.475\textwidth]{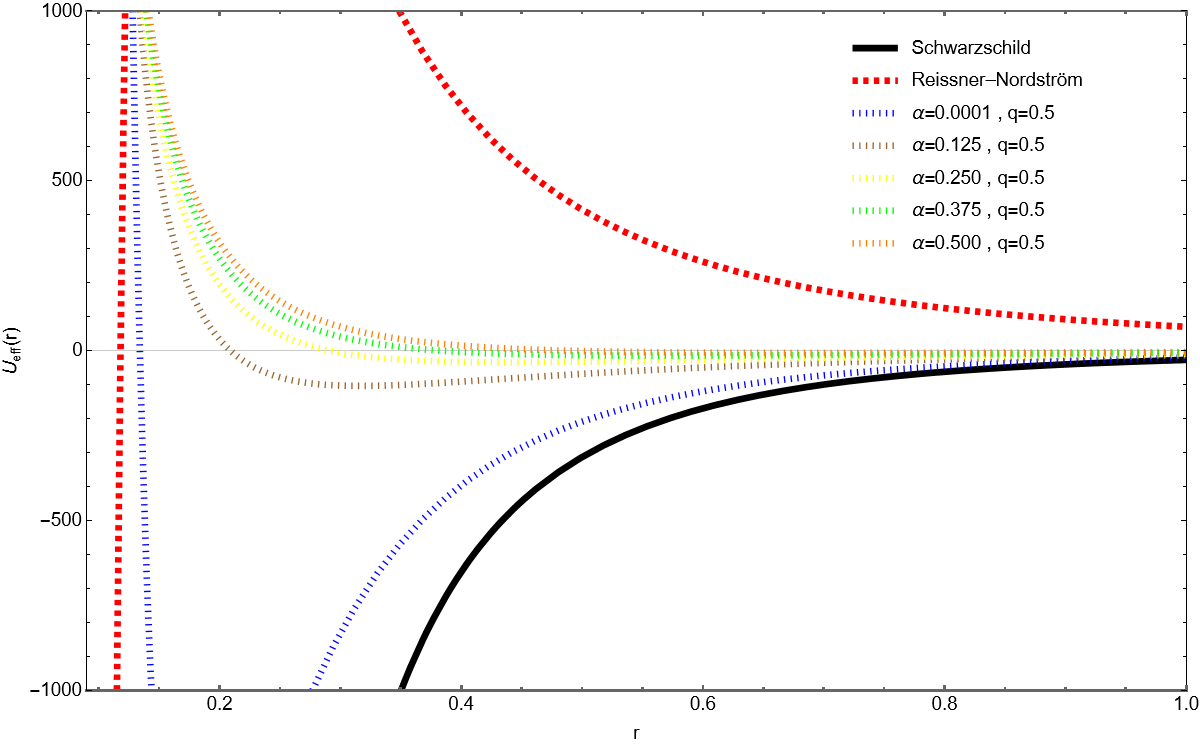}}
\,\,\,
\subfloat[\label{Fig1c} for $\alpha=0.125$]{\includegraphics[width=0.475\textwidth]{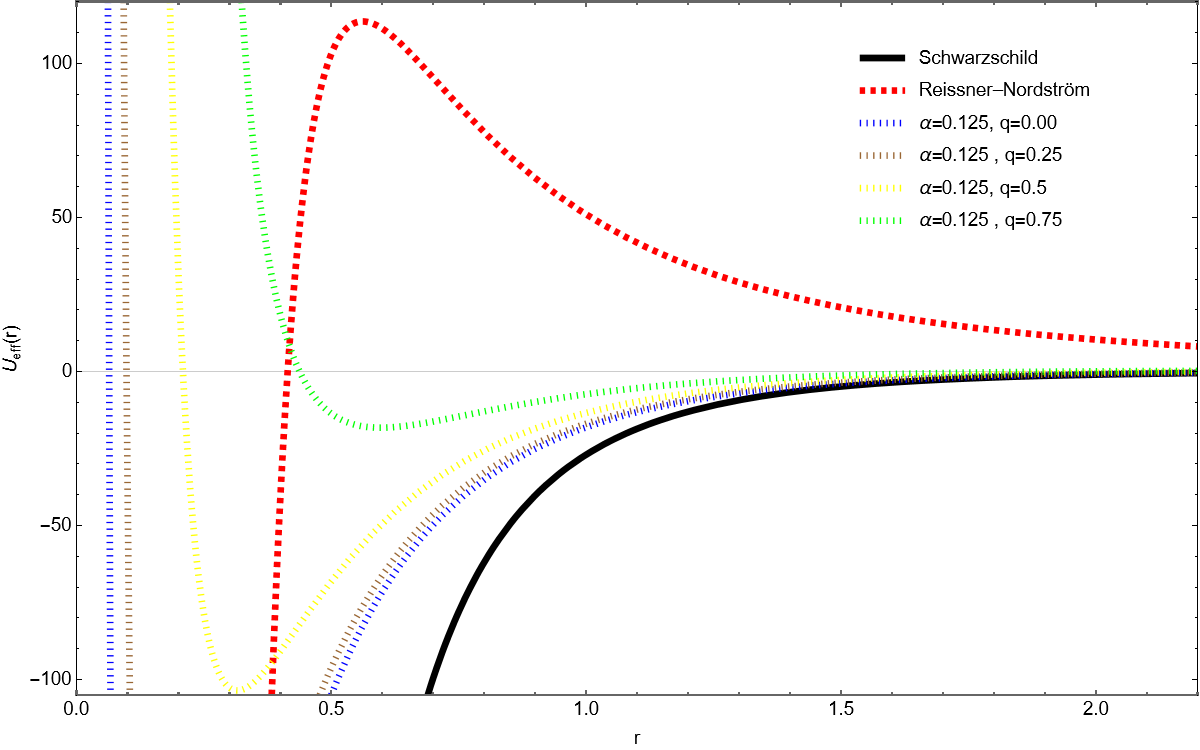}}
\caption{\label{Fig5}\small{\emph{The radial variation of the effective potential of the  Extreme Chgarged Compact Object in 4D Einstein-Gauss-Bonnet gravity for various values of $\alpha$ and $q$, where we set $L = 5$ and $E = K = M = 1$.
}}}\label{Fig1}
\end{figure}

Figure.\ref{Fig1} displays the effective potential for a Extreme Compact Charged Object  in four-dimensional Einstein-Gauss-Bonnet gravity as a function of the radial coordinate \( r \), for various values of the electric charge \( q \) and the Gauss-Bonnet coupling parameter \( \alpha \). The maximum of the effective potential for each pair of \( q \) and \( \alpha \) corresponds to the photon sphere radius, denoted by \( r_0 \). As shown in Figs.\ref{Fig1a} and \ref{Fig1b}, increasing the Gauss-Bonnet coupling parameter \( \alpha \)  (that is to say, enhancing the stringy effects), while keeping the electric charge \( q \) fixed, leads to an increment  in the effective potential  and this means more effectiveness of the gravitational effect. Furthermore, as illustrated in Fig.\ref{Fig1c}, increasing the electric charge \( q \), for a fixed value of \( \alpha \), also results in an increase in the effective potential. This is expected, as usual, for electromagnetic systems.

\subsection{Shadow geometry}\label{MTP}
A black hole shadow is a two-dimensional image in the sky where light paths are deflected by the black hole's strong gravitational field and trapped by the black hole instead of reaching the observer. In fact, the black hole shadow is the boundary between photons escaping the strong gravitational field and photons trapped in unstable photonic orbits and ultimately trapped by the black hole. This boundary is surrounded by a photon ring. Studying the shadow is a key tool for testing general relativity in the strong field regime, estimating the fundamental properties of black holes, and investigating alternative models of gravity. Data from the Event Horizon Telescope (EHT), first released in 2019 for M87* and in 2022 for Sagittarius A*, have provided excellent evidence for studying the shadows of supermassive black holes.

To proceed, we focous on the profile and size of the shadows of black holes in the background of the spacetime metric  defined by Eq.~\eqref{rrr} in arbitrary dimensions. We firstly define two impact parameters \(\xi\) and \(\eta\), in terms of the constants of the motion \(E\), \(L\), and \(K\). These parameters describe the properties of photons' orbits  in the vicinity of the  black hole and are given by:
\begin{equation}
\xi = \frac{L}{E} \ , \quad \eta = \frac{K}{E^2}  \ ,
\label{eq:impact_params}
\end{equation}

Using these definitions, the effective potential \(V_{\text{eff}}\) and the function \(R\) can be rewritten in terms of \(\xi\) and \(\eta\) as follows:
\begin{equation}
V_{\text{eff}} = E^2 \left[ \frac{f(r)}{r^2} \left( \eta + \xi^2 \right) - 1 \right],\label{mar10}
\end{equation}
\begin{equation}
R = E^2 \left[ r^4  - r^2 f(r) \left( \eta + \xi^2 \right) \right],\label{mar11}
\end{equation}
By substituting  Eqs.~\eqref{mar10} and \eqref{mar11} into Eq.~\eqref{eq:mar}, we obtain the following relation involving the impact parameters \(\xi\) and \(\eta\):
\begin{equation}
\eta + \xi^{2} = \frac{4 r_0^{2}}{r_0 f'(r_0) + 2 f(r_0)} \ .
\label{eq:impact_parameters}
\end{equation}

Here \( r_0 \) is the photon sphere radius (a length); while the combination \( \eta + \xi^2 \) is a length-squared quantity. We need these impact  parameters to study the motion of photons and to describe  the shadow of a black hole. In the astrophysical observation, we can use the celestial coordinates \(\lambda\) and \(\psi\) to describe the apparent shape of BH shadow as observed by a remote observer  \cite{Vazquez:2003zm}. These coordinates are 
\begin{figure}[H]
    \centering
    \includegraphics[width=0.5\textwidth]{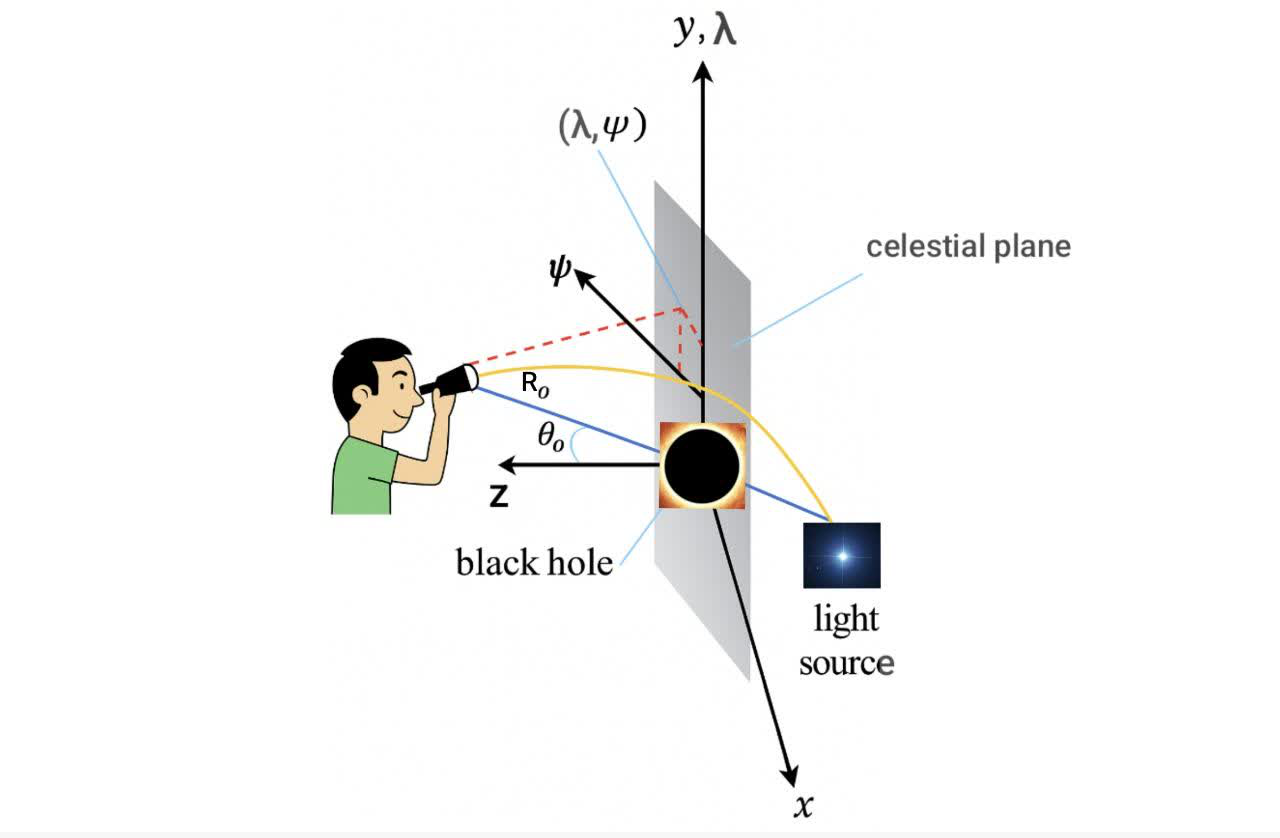}
    \caption{The celestial coordinates on the distant observer's sky are shown. 
The observer's position is $(R_o, \tilde{\theta}_o)$, 
and $(\lambda, \psi)$ gives the image's apparent position.}
    \label{fig:plot18}
\end{figure}
\begin{equation}
\lambda = \lim_{R_o \to \infty} \left( \frac{R_o^2 P(\theta_{D-2})}{P(t)} \right),
\quad
\psi = \lim_{R_o \to \infty} \left( \frac{R_o^2 P(\theta_i)}{P(t)} \right),
\end{equation}
where \(P(t)\), \(P(\theta_{D-2})\) and \(P(\theta_i)\) are the momenta in the observer's frame, and \(R_o\) is the radial distance of the observer from the black hole. Because the metric spacetime we are studying is four-dimensional,  we set \( D = 4 \).  On the equatorial plane, these translate to \(\psi = \pm \sqrt{\eta }\) and \(\lambda = -\xi \). 

Importantly, the squared shadow radius in celestial coordinates, \(r_s\), is given by
\begin{equation}
r_s^2 \equiv \eta + \xi^2 = \lambda^2 + \psi^2,\label{eq}
\end{equation}

which describes the relation between the impact parameters in terms of the observable shadow geometry. For static black holes (without rotation), this leads to a perfect circle of radius \(r_s\).

Here,  we study the geometric structure of the shadow of a four-dimensional Einstein-Gauss-Bonnet Extreme Compact Charged Object photographed on the celestial sphere.We start by gathering the involved quantities: \( r_{eh} \), \( r_0 \), and \( \sqrt{\eta + \xi^2} \), which represent the event horizon radius, the photon sphere radius, and the shadow radius (according to Eq.~\eqref{eq}), respectively. The value of the photon sphere radius \( r_{\text{o}} \) is obtained by substituting  Eq.~\eqref{eq:F_r} into  Eq.~\eqref{eq:mar2}. The  radius of the shadow in celestial coordinates  is obtained through Eq.~\eqref{eq} after applying Eq.~\eqref{eq:F_r} into Eq.~\eqref{eq:impact_parameters}. Numerical values of \( r_{eh} \) , \( r_{S} \)  and \( r_0 \) for various combinations of the Gauss--Bonnet coupling constant \( \alpha \) and electric charge \(q \) are summarized in Table~\ref{Table1}.

\begin{table}[H]
  \centering
  \caption{\label{Table1}\small{\emph{Values of \( r_{eh} \),\( r_0 \) and \( r_{S} \) for different values of \( \alpha \) and \( q \).}}}
  {\renewcommand{\arraystretch}{1.3}
  \begin{tabular}{|c||c|c|c||c|c|c||c|c|c||c|c|c|}
  \hline
  & \multicolumn{3}{|c||}{$q=0.00$} & \multicolumn{3}{|c||}{$q=0.25$} & \multicolumn{3}{|c||}{$q=0.5$} & \multicolumn{3}{|c||}{$q=0.75$}\\ \cline{2-13}
  $\alpha$ & $r_{_{eh}}$ & $r_{_{0}}$ & $r_{_{S}}$& $r_{_{eh}}$ & $r_{_{0}}$ & $r_{_{S}}$& $r_{_{eh}}$ & $r_{_{0}}$ & $r_{_{S}}$& $r_{_{eh}}$ & $r_{_{0}}$ & $r_{_{S}}$\\\hline\hline
  $0.00001$ & $1.999$ & $2.991$ &$5.196$ & $1.969$ &$2.958$ & $5.141$ &$1.866$ & $2.823$ &$4.968$ & $1.661  $ &$2.560$ & $4.638$  \\\hline
  $0.125$ & $1.935$ & $2.943$ &$5.147$ & $1.901$ &$2.899$ & $5.090$ &$1.790$ & $2.756$ &$4.911$ & $1.559$ &$2.474$ & $4.565$  \\\hline
  $0.250$ & $ 1.866$ & $2.882$ &$5.095$ & $1.829$ &$2.835$ & $5.037$ &$1.707$ & $2.685$ &$4.850$ & $1.433$ &$2.375$ & $4.484$     \\
  \hline
 $0.375$ & $1.790$ & $2.817$ &$5.040$ & $1.750$ &$2.767 $ & $4.979$ &$1.612$ & $2.605$ &$4.784$ & $1.250$ &$2.257$ & $4.391$     \\
  \hline
 $0.500$ & $1.707$ & $2.746$ &$4.982$ & $1.661$ &$2.694$ & $4.919$ &$1.500$ & $2.517$ &$4.713$ & $ \text{NotReal}$ &$2.106$ &$4.281$    \\
  \hline

  \end{tabular}}
\end{table}
\FloatBarrier
\begin{figure}[H]
\centering
\subfloat[for $\alpha=0.125$]{%
  \includegraphics[width=0.475\textwidth]{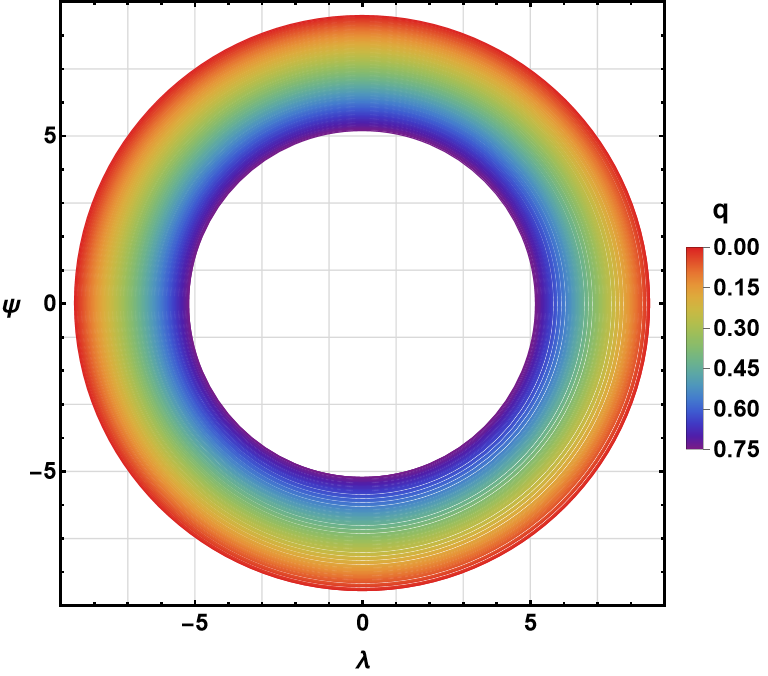}%
  \label{Fig2a}
}
\,\,\,
\subfloat[for $\alpha=0.500$]{%
  \includegraphics[width=0.475\textwidth]{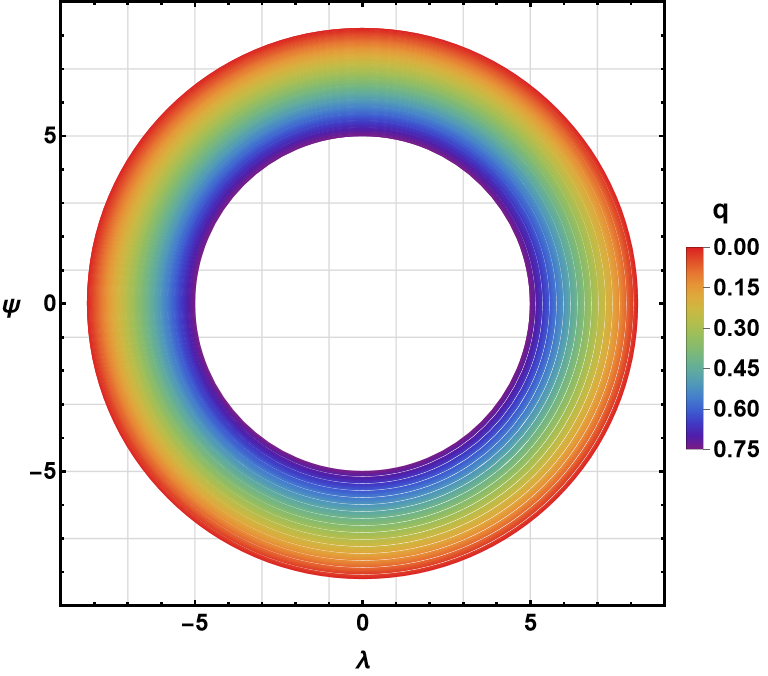}%
  \label{Fig2b}
}
\caption{\label{Fig5}\small{\emph{Geometrical shape of the shadow of the  Extreme chgarged compact object in 4D Einstein-Gauss-Bonnet gravity (with  $M = 1$).}}}\label{Fig2}
\end{figure}

In Figure.\ref{Fig2}, we illustrate the shadow geometry of the Extreme Compact Charged Object  in 4D Einstein-Gauss-Bonnet gravity for different values of $q$. Fig.\ref{Fig2a} is for $\alpha = 0.125$ and Fig.~\ref{Fig2b} is for $\alpha = 0.500$. By studying each of the figures separately, we  see that for a fixed value of $\alpha$, when the electric charge increases, the shadow radius gets  smaller. Moreover, comparing Fig.\ref{Fig2a} and  Fig.\ref{Fig2b}, we observe that by  increasing $\alpha$, the shadow radius of the  Extreme Compact Charged Object    becomes smaller in the 4D Einstein-Gauss-Bonnet gravity. It is expected that in the future and in the next generation of the EHT, traces of these effects will be observed by measuring the shadow radius of a greater number of supermassive black holes.

\begin{figure}[H]
\centering
\subfloat[for $q=0.00$]{%
  \includegraphics[width=0.475\textwidth]{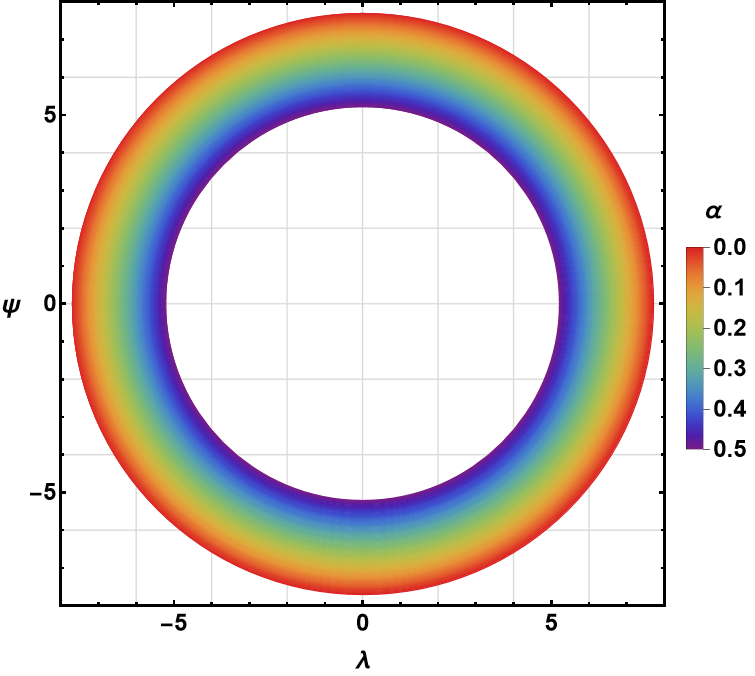}%
  \label{Fig3a}
}
\,\,\,
\subfloat[for $q=0.25$]{%
  \includegraphics[width=0.475\textwidth]{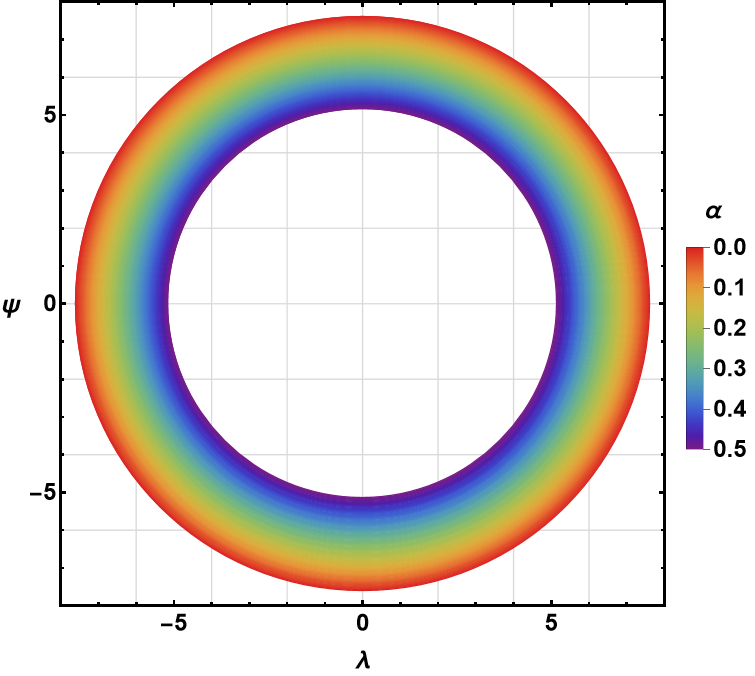}%
  \label{Fig3b}
}
\,\,\,
\subfloat[for $q=0.5$]{%
  \includegraphics[width=0.475\textwidth]{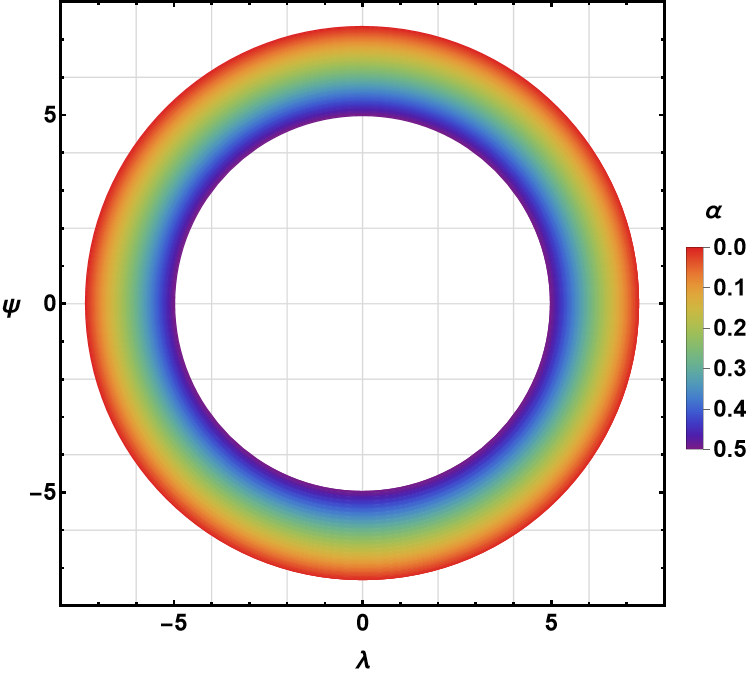}%
  \label{Fig3c}
}
\,\,\,
\subfloat[for $q=0.75$]{%
  \includegraphics[width=0.475\textwidth]{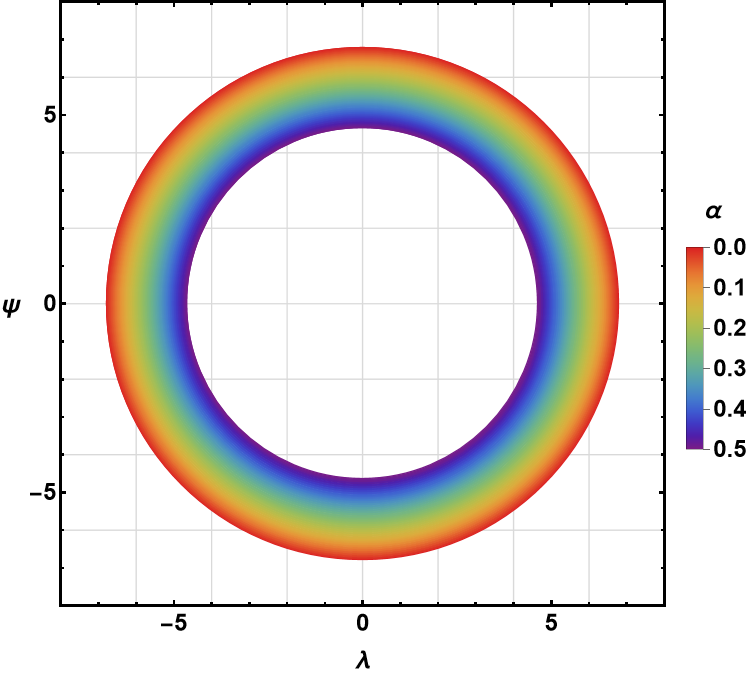}%
  \label{Fig3d}
}
\caption{\label{Fig3}\small{\emph{Geometrical shape of the shadow of the  Extreme Chgarged Compact Cbject in 4D Einstein-Gauss-Bonnet gravity( with $M = 1$).}}}
\end{figure}
For the observational aspect of the  shadow of the Extreme Compact Charged Object  in 4D EGB gravity, we can use the given data in Table \ref{Table1} to draw the shadow of this  Extreme Compact Charged Object  for various values of the parameters \(q\) and \(\alpha\). In drawing  Fig.\ref{Fig3}, for each figure the value of \(q\) is fixed. Looking at all these plots, there is the general feature that increasing the  parameter \(\alpha\) at fixed \(q\), diminishes the shadow radius of the  Extreme Compact Charged Object. Also, from Figures.\ref{Fig3a}, \ref{Fig3b}, \ref{Fig3c}, and \ref{Fig3d}, we see that larger electric  charge \(q\) leads to smaller shadow size.

\subsection{Energy emission rate}
It is known that black holes can radiate through a phenomenon called Hawking radiation, and at very high energies, the absorption cross-section generally oscillates around a limiting value $\sigma_{\lim}$. An important point is that for an observer located very far away from the black hole (or Extreme Compact Charged Object), this absorption cross-section advances toward the black hole shadow  \cite{Wei:2013kza}. It is possible to demonstrate that $\sigma_{\lim}$ is roughly equivalent to the area of the photon sphere, which may be expressed as follows in arbitrary dimension~\cite{Wei:2013kza}:
\begin{equation}
\sigma_{\lim} \approx \left(\frac{\pi^{\frac{D-2}{2}}}{\Gamma\left( \frac{D}{2} \right)}\right)r_s^{D-2},
\label{eq:sigma_lim}
\end{equation}
where $r_s$ is the radius of the shadow.
The form of the energy emission rate for a  Extreme chgarged compact object is given as follows :
\begin{equation}
\frac{d^2E(\varpi)}{d\varpi\,dt} = \frac{2\pi^2 \sigma_{\lim}}{e^{\varpi/T} - 1} \, \varpi^{D-1},
\label{eq:energy_emission}
\end{equation}
where $\varpi$ is the emission frequency $T$ is the Hawking temperature given as:
\begin{equation}
T = \frac{1}{4 \pi r_{\text{eh}}}\ .
\label{eq:HawkingTemp}
\end{equation}
To study the Hawking temperature of the Extreme Compact Charged Object in four-dimensional Einstein–Gauss–Bonnet gravity for different values of the parameters $\alpha$ and $q$, we consider the values of $r_{\text{eh}}$ reported in Table~\ref{Table1}. It is obvious that  the number of spacetime dimensions (D) is four. To obtain the values of $\sigma_{\lim}$ for different values of $\alpha$ and $q$, we substitute the quantity $\sqrt{\eta + \xi^2}$ into Eq.~\eqref{eq:impact_parameters}. Then, by plugging the Hawking temperature and $\sigma_{\lim}$ into Eq.~\eqref{eq:energy_emission}, we obtain the energy emission rate for a  Extreme Compact Charged Object     in Einstein–Gauss–Bonnet gravity in four dimensions as a function of frequency for various values of $\alpha$ and $q$.
\begin{figure}[H]
\centering
\subfloat[\label{Fig4a} for q=0.00]{\includegraphics[width=0.475\textwidth]{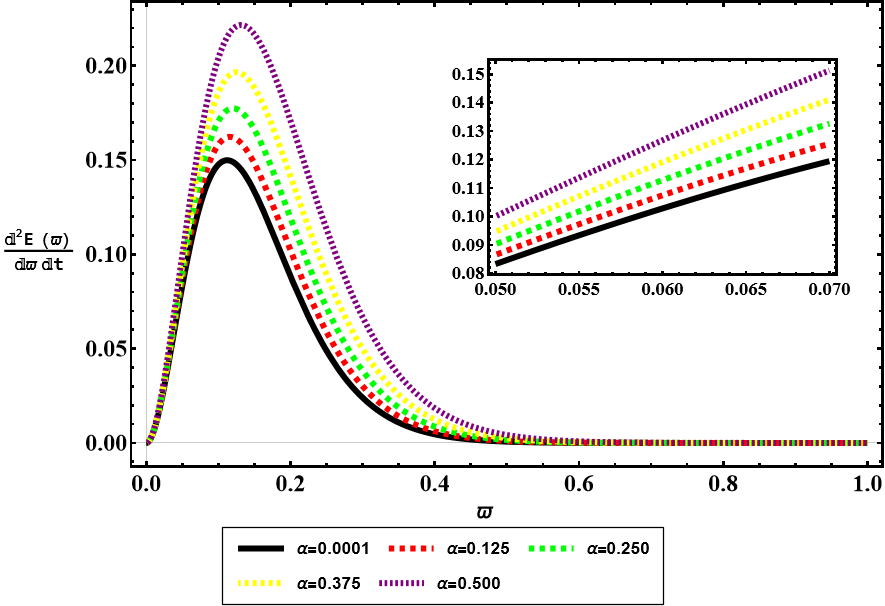}}
\,\,\
\subfloat[\label{Fig4b} for q=0.5]{\includegraphics[width=0.475\textwidth]{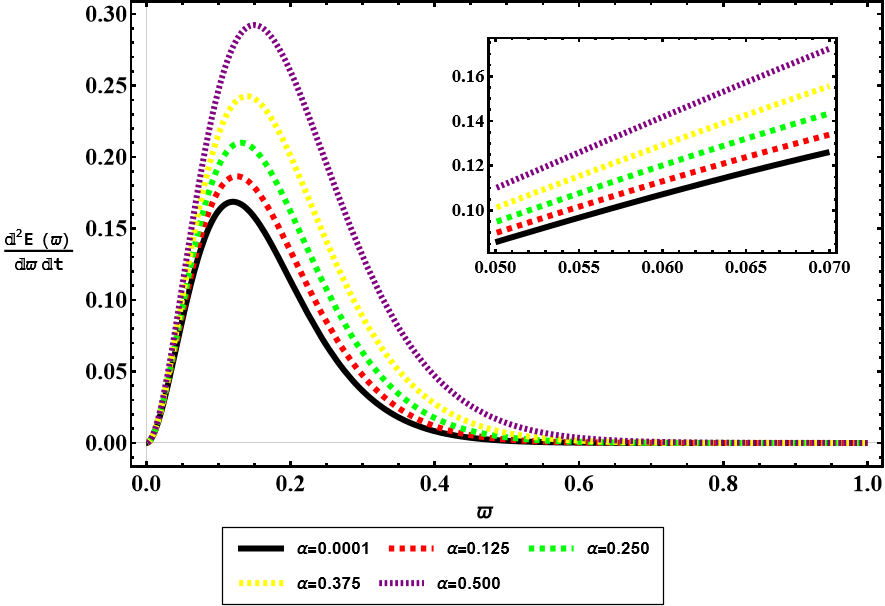}}
\,\,\
\subfloat[\label{Fig4c} for $\alpha=0.0001$]{\includegraphics[width=0.475\textwidth]{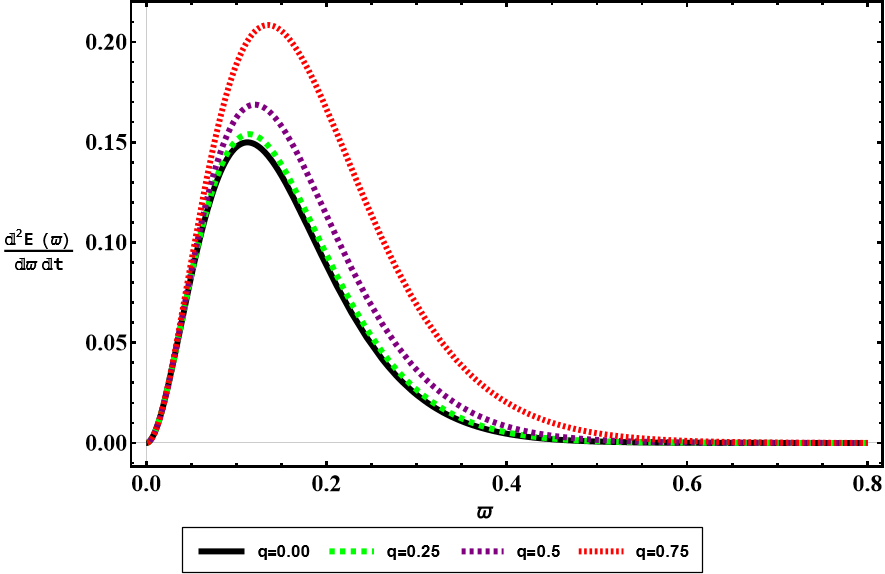}}
\,\,\
\subfloat[\label{Fig4d} for $\alpha=0.375$]{\includegraphics[width=0.475\textwidth]{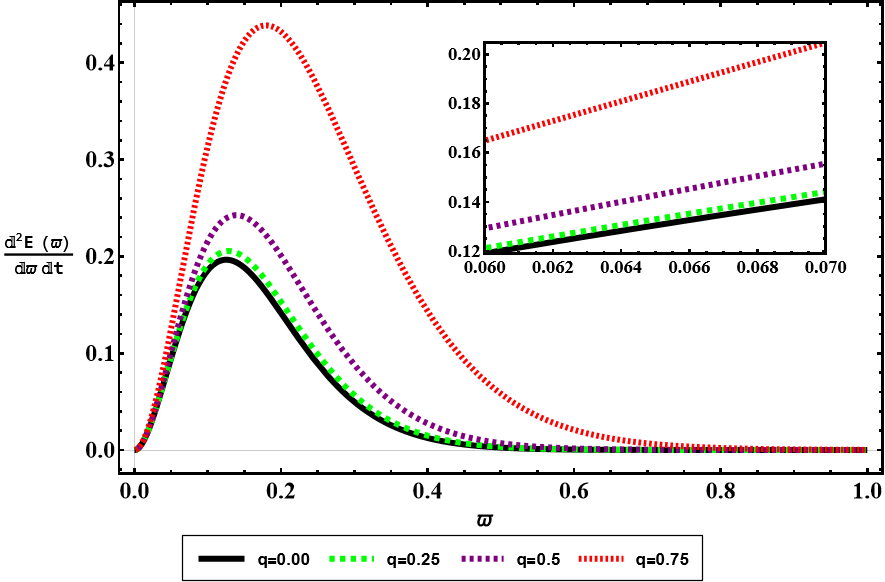}}
\caption{\label{Fig4}\small{\emph{The energy emission rate as a function of $\varpi$ for different values of $q$ and $\alpha$ for a  Extreme Compact Charged Object     in 4D EGB gravity.}}}
\end{figure}

Figure.\ref{Fig4} displays the energy emission rate plotted as a function of the frequency for the Extreme chgarged compact object in 4D EGB gravity. By looking at Figs. \ref{Fig4a} and \ref{Fig4b}, we understand that the energy emission rate for the  Extreme chgarged compact object increases with different values of \(\alpha\) at a fixed value of \(q\). From this point of view, this means that the larger the value of \(\alpha\), the faster the Extreme Compact Charged Object evaporates.  By examining Figs. \ref{Fig4c} and .\ref{Fig4d}, we learn that, for a larger \(q\), evaporation of the Extreme Compact Charged Object in 4D Einstein–Gauss–Bonnet gravity becomes faster. This means that  4D EGB charged Extreme Compact Charged Objects evaporate  more quickly than the chaegeless  counterparts.

\subsection{Deflection Angle}
In this section, we aim to study the bending angle of light around the  Extreme chgarged compact object in the context of the four-dimensional spacetime in Einstein-Gauss-Bonnet gravity. To do this end, we use the Gauss-Bonnet theorem \cite{Arakida:2017hrm}. We start by calculating the optical metric restricted to the equatorial plane with \(\theta = \pi/2\) in the spacetime described by line element~\eqref{eq:metric}. Then, for null geodesics on this plane, with \( ds^2 = 0 \), we obtain the optical metric as follows:

\begin{equation}
dt^2 = \frac{dr^2}{f ^2(r)} + \frac{r^2}{f(r)} d\phi^2,\label{eq:MAR3}
\end{equation}
For this optical metric, we can calculate the Gaussian optical curvature \(\mathcal{K} = \frac{\bar{R}}{2}\) in which \(\bar{R}\) is the Ricci scalar of the metric \eqref{eq:MAR3} 
\begin{equation}
\mathcal{K} = \frac{2 f'(r) f ^2(r) - f'^2(r) f(r) \, r}{4 f(r) \, r} \ .
\end{equation}
In order to calculate the deflection angle, one should consider a non-singular manifold \(\tilde{D}_R\) with a geometrical size \(\tilde{R}\) to employ the Gauss-Bonnet theorem, so that \cite{Arakida:2017hrm}
\begin{equation}
\iint_{\tilde{D}_R}\mathcal{K} \, dS + \oint_{\partial \tilde{D}_R} \kappa \, dt + \sum_i \phi_i = 2\pi \, \zeta(\tilde{D}_R),
\end{equation}
where \( dS = \sqrt{\tilde{g}}\, dr\, d\varphi \) and \( dt \) are the surface element and squared  line element of the optical metric \eqref{eq:MAR3}, respectively. \(\tilde{g}\) is the determinant of the optical metric, \(\kappa \) denotes the geodesic curvature of \(\tilde{D}_R\), and \(\phi_i\) is the jump (exterior) angle at the \(i\)-th vertex, and also, \(\zeta(\tilde{D}_R)\) is the Euler characteristic number of \(\tilde{D}_R\). One can set \(\zeta(\tilde{D}_R) = 1\). Then, considering a smooth curve \( y \), which has the tangent vector \(\dot{y}\) and acceleration vector \(\ddot{y}\), the geodesic curvature \(\kappa  \) of \( y \) can be defined as follows where the unit speed condition \(\tilde{g}(\dot{y}, \dot{y}) = 1\) is employed:
\begin{equation}
\kappa = \tilde{g} \big( \nabla_{\dot{y}} \dot{y}, \ddot{y} \big),
\end{equation}
which is a measure of deviation of $y$ from being a geodesic. In the limit $\tilde{R} \to \infty$, two jump angles $\varphi_s$ (of source) and $\varphi_O$ (of observer) will become $\pi/2$, i.e., $\varphi_s + \varphi_O\to \pi$. Considering $C_{\tilde{R}} := r(\varphi)$, we have $\kappa(C_{\tilde{R}}) = \left| \nabla_{\dot{C}_{\tilde{R}}} \dot{C}_{\tilde{R}} \right|_{\tilde{R} \to \infty} \to 1/\tilde{R}$, and therefore, we find $\lim_{\tilde{R} \to \infty} dt = \tilde{R} \, d\varphi$. Hence, $\kappa(C_{\tilde{R}}) \, dt = d\varphi$. Consequently, the Gauss-Bonnet theorem will reduce to the following form:
\begin{equation}
\iint_{D_{\tilde{R}}}\mathcal{K}\, dS + \int_{C_{\tilde{R}}}\kappa \, dt \Big|_{\tilde{R} \to \infty} = 
\iint_{D_{\infty}} \mathcal{K} \, dS + \int_{0}^{\pi + \Theta} d\varphi = \pi   \ .
\label{eq:47}
\end{equation}

Thus, following equation for calculating the deflection angle (see Refs. \cite{Arakida:2017hrm} and references therein), we have
\begin{equation}
\Phi= \pi - \int_{0}^{\pi + \Theta} d\varphi = - \int_{0}^{\pi} \int_{\frac{\xi}{\sin \varphi}}^{\infty}\,\mathcal{K} \, dS
\label{eq:deflection}  \ .
\end{equation}
Now, by substituting the metric function under study into equations \eqref{eq:47} and \eqref{eq:deflection}, we can calculate the Gaussian optical curvature for a  Extreme Compact Charged Object    in four-dimensional Einstein-Gauss-Bonnet gravity, which is expressed as follows:
\begin{equation}
\mathcal{K} = \frac{1 - \sqrt{1 + 4 \alpha \left( \frac{2 m r - q^{2}}{r^{4}} \right)}}{2 \alpha}  \ . 
\label{eq:K}
\end{equation}
Moreover, the optical metric surface element (equation \eqref{eq:MAR3}) for a  Extreme chgarged compact object in four-dimensional Einstein-Gauss-Bonnet gravity (equation \eqref{rrr}), according to the metric coefficient, is given by the following expression.For large distances $r$, the metric function $f(r) \to 1$, 
so the surface element simplifies to its leading-order term, which is sufficient for calculating the asymptotic deflection angle:
\begin{equation}
dS = \sqrt{\bar{g}} \, dr d\varphi = \frac{r}{{f(r)} \sqrt{f(r)}} \, dt d\varphi \approx r \, dr d\varphi
\label{eq:dS},
\end{equation}
The deflection angle of the Extreme Compact Charged Object in four-dimensional Einstein-Gauss-Bonnet gravity is given as follows:
\begin{equation}
\Phi= - \int_{0}^{\pi} \int_{\frac{\xi}{\sin \varphi}}^{\infty}\,\mathcal{K} \, dS = - \int_{0}^{\pi} \int_{\frac{\xi}{\sin \varphi}}^{\infty}\,\frac{1 - \sqrt{1 + 4 \alpha \left( \frac{2 m r - q^{2}}{r^{4}} \right)}}{2 \alpha}\, r \, dr \, d\varphi
\label{eq:Theta}
\end{equation}
After a lengthy calculation, the integral in Eq. \ref{eq:Theta} evaluates to the following expression in terms of a hypergeometric function:
\[
 = \frac{\pi \left( 
\sqrt{-q^2 + 2 m r} \, 
\sqrt{(-q^2 + 2 m r) \alpha} \left(12 b^4 + 5 (q^2 - 2 m r) \alpha \right)
- 
\frac{3 b^8 \left( \frac{(-q^2 + 2 m r) \alpha}{b^4} \right)^{3/2} 
\sqrt{\frac{b^4}{-q^2 \alpha + 2 m r \alpha}} 
\, {}_{3}F_{2} \left(\frac{1}{2}, \frac{3}{4}, \frac{5}{4}; \frac{3}{2}, 2; \frac{4 (q^2 - 2 m r) \alpha}{b^4} \right)
}{\sqrt{\alpha}}
\right)}
{12 b^6 \sqrt{\alpha}} ,\label{eq:b}
\]
Where \({}_{3}F_{2}\) shows the hypergeometric function . By substituting \(-q^{2} + 2 m r = \Psi\), equation \eqref{eq:Theta} simplifies to:
\begin{equation}\label{eq:b2}
\Phi= \frac{\pi \left( 
\sqrt{\Psi} \, 
\sqrt{\Psi \alpha} \left(12 b^4 + 5 (-\Psi) \alpha \right)
- 
\frac{3 b^8 \left( \frac{\Psi \alpha}{b^4} \right)^{3/2} 
\sqrt{\frac{b^4}{\Psi \alpha}} 
\, {}_{3}F_{2} \left(\frac{1}{2}, \frac{3}{4}, \frac{5}{4}; \frac{3}{2}, 2; \frac{4 (-\Psi) \alpha}{b^4} \right)
}{\sqrt{\alpha}}
\right)}
{12 b^6 \sqrt{\alpha}} .
\end{equation}

\begin{figure}[H]
\centering
\subfloat[\label{Fig5a} for $\alpha=0.250$]{\includegraphics[width=0.475\textwidth]{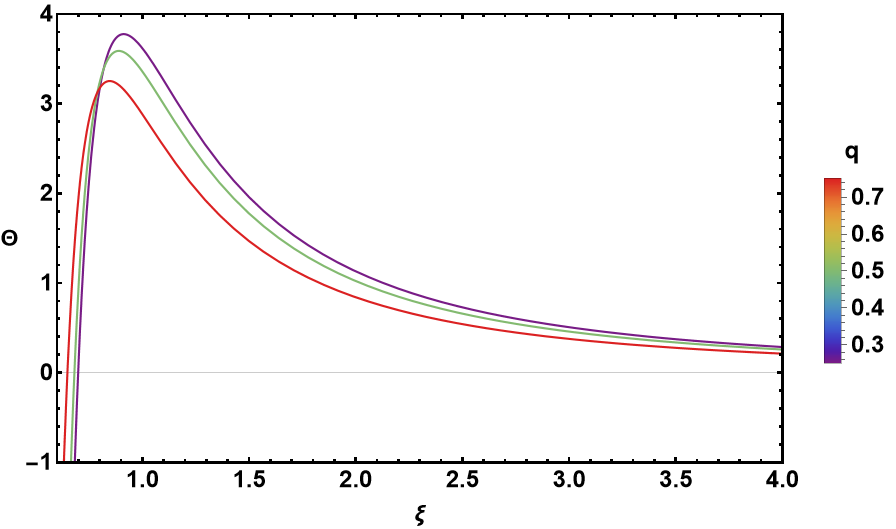}}
\,\,\

\subfloat[\label{Fig5b} for q=0.25]{\includegraphics[width=0.475\textwidth]{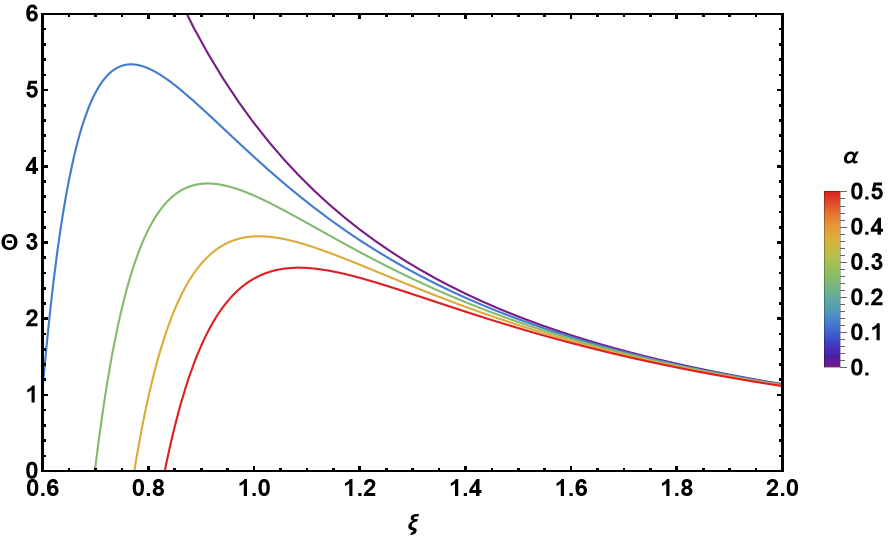}}
\,\,\
\subfloat[\label{Fig5c} for q=0.75]{\includegraphics[width=0.475\textwidth]{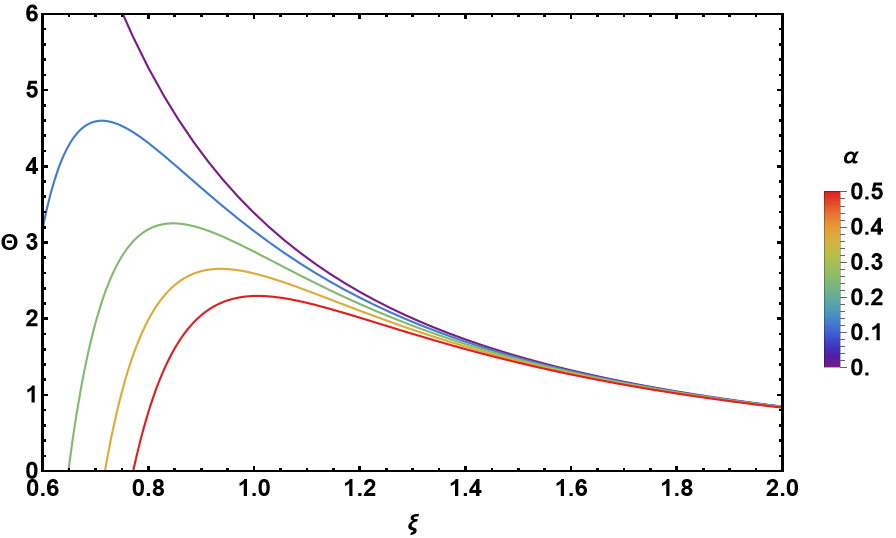}}
\caption{\label{Fig5}\small{\emph{The behavior of the deflection angle of the Extreme Compact Charged Object  in 4-dimensional Einstein-Gauss-Bonnet gravity as a function of $\xi$ for different values of $\alpha$ and $q$.}}}
\end{figure}

In Figure.~\ref{Fig5}, we present the deflection angle of the Extreme chgarged compact object in the four-dimensional Einstein-Gauss-Bonnet gravity. The bending angle is depicted in Fig.~\ref{Fig5a} as a function of \(q\) for a fixed value of \(\alpha=0.250\). Figs.~\ref{Fig5b} and \ref{Fig5c} reveal the bending angle as a function of \(\alpha\) for fixed values \(q=0.25\) as well as \(q=0.75\).  As is seen in Fig.~\ref{Fig5a}, for  a smaller values  of the impact parameter \(\xi\),  the deflection angle of the  Extreme Compact Charged Object  increases. Also from Figure~\ref{Fig5a}, it is observed that, for a fixed value of the parameter \(\alpha\), the deflection angle of the dark compact  object can be diminished by increasing the electric charge. In Figs.~\ref{Fig5b}  and \ref{Fig5c}, we notice that for a fixed magnitude of charge \((q)\), the increment  of the parameter \((\alpha)\) causes the deflection angle to decrease. From Figs.~\ref{Fig5b} and \ref{Fig5c} we notice that with an increment  in the electric charge, the gradient of the deflection angle decreases. The increase of the electric charge leads to a decrease in the deflection of light, and moreover, the reduction in the slope of the deflection angle with increasing charge indicates that not only the overall deflection decreases, but also the sensitivity of the deflection angle to variations of the other parameter, namely $\alpha$ (the Gauss-Bonnet parameter), is reduced. Both the electric charge and the Gauss-Bonnet parameter cause a decrease in the deflection angle, but the effect of the charge is dominant.

\section{Constraints from EHT observations}\label{ARMECCO}

In this section, our goal is to compare the radius obtained for the shadow of the  Extreme compact charged  object  in the four-dimensional Einstein–Gauss–Bonnet gravity with the size of the supermassive black hole M87*  shadow recorded by the Event Horizon Telescope \cite{EventHorizonTelescope:2021dqv}. In the EHT data, the radius of the black hole's shadow is estimated to be \( 4.31 \leq R_{s, \mathrm{M87}^*} \leq 6.08 \). Using this constraint, we can impose new constraint on the Gauss–Bonnet parameter $\alpha$.
\begin{figure}[H]
\centering
\includegraphics[width=0.7\textwidth]{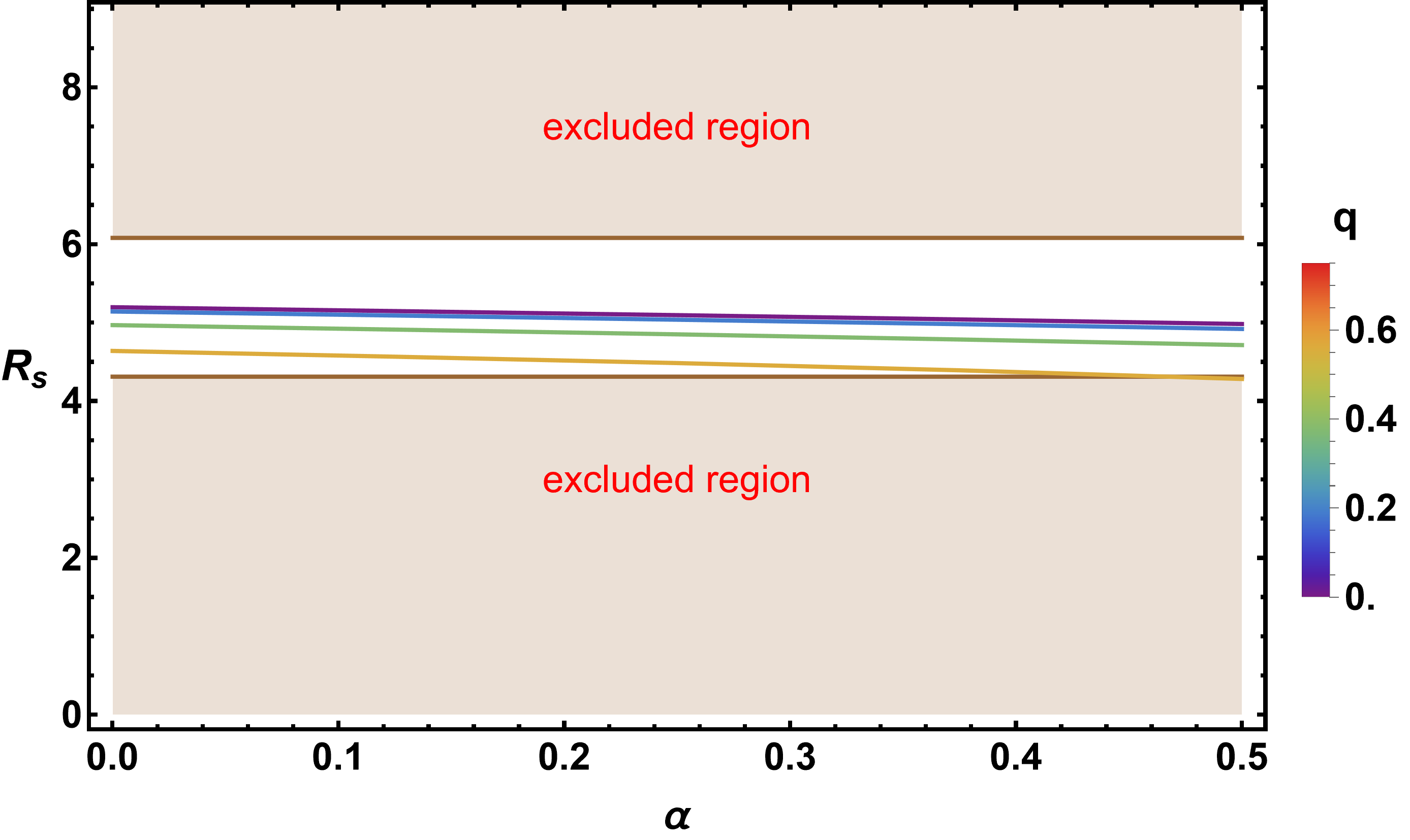}
\caption{\small{\emph{The shadow radius of the  Extreme chgarged compact object in four-dimensional Einstein--Gauss--Bonnet gravity, compared to the shadow size of \( M87^{*} \) captured by the EHT, versus the parameter \(\alpha\). The colored area are the excluded regions, which are  inconsistent with the EHT data, while the white region corresponds to the values consistent with EHT observations.
}}}
\label{fig6}
\end{figure}
Figure~\ref{fig6} shows the behavior of the  Extreme Compact Charged Object shadow in four-dimensional Einstein--Gauss--Bonnet gravity compared to the shadow recorded by the EHT for M87$^*$ at the 1-$\sigma$ (i.e., 68\%) confidence level in terms of the Gauss--Bonnet coupling constant~$\alpha$. From the figure, we observe that almost all shadow radii of the  Extreme Compact Charged Object    in four-dimensional Einstein-Gauss-Bonnet gravity are consistent with the EHT observational data at the 68\% confidence level. It is only observed that by increasing the electric charge \( q \), the shadow radius of the EGB Extreme Compact Charged Object gradually deviates from the EHT data. Moreover, the main goal of all calculations and analyses was to constrain the Gauss-Bonnet parameter \( \alpha \). It is evident that within the range \( 0 \leq \alpha < 0.500 \), the shadow radius of the Extreme Compact Charged Object shows very good agreement with the EHT observations. Initially, a larger range of $\alpha$ was considered, but for values of $\alpha$ greater than 0.5, the black hole did not have an event horizon and became singular. Therefore, the range of $\alpha$ was set between 0 and 0.5. This constrained range for $\alpha$ is consistent with, and provides a valuable independent check on, bounds derived from other astrophysical phenomena and theoretical consistency within the 4D EGB framework, which typically suggest $\alpha$ is positive and of order unity.

\section{Summary and Conclusions}\label{SaC}
In this study, our motivation was to investigate the behavior of a  Extreme Compact Charged Object in Einstein-Gauss-Bonnet gravity in four dimensions. We investigated the shadow behavior and the deflection angle of the corresponding  Extreme Compact Charged Object  in EGB and tried to impose constraints on the Gauss-Bonnet coupling constant $\alpha$ by matching our study with EHT observational data. First, we calculated the null geodesics and effective potentials using the Hamilton-Jacobi approach and the Carter method. Then, we used the celestial coordinates to determine the shape of the shadow of the 4D EGB   Extreme Compact Charged Object    on the observer's sky.

Then, the shadow behavior, deflection angle, and energy emission rate of a 4D EGB  Extreme Compact Charged Object were investigated. We constrained the Gauss-Bonnet coupling constant $\alpha$ by comparing the shadow size of M87* obtained from EHT observations with the shadow radius of the 4D EGB  Extreme Compact Charged Object    towards faster evaporation.

For the four-dimensional EGB  Extreme Compact Charged Object, we found that increasing the Gauss-Bonnet parameter $\alpha$ leads to a smaller shadow size. Also, we see that the energy emission rate of the EGB  Extreme chgarged compact object increases with increasing $\alpha$. Therefore, we found that the Gauss-Bonnet term can significantly affect the evaporation of the Extreme Compact Charged Object.

Then, using the Gauss-Bonnet theorem, we determined  the deflection angle and found that the deflection angle of the EGB  Extreme Compact Charged Object  decreases with increasing $\alpha$. Finally, we found that the shadow of a four-dimensional EGB   Extreme Compact Charged Object with Gauss-Bonnet coupling constant $\alpha$ is in agreement with the EHT data.
In summary, we conclude that the shadow of the four-dimensional Einstein-Gauss-Bonnet  Extreme Compact Charged Object   , for values of $\alpha$ in the range $0 \leq \alpha < 0.5$, is in very good agreement with the supermassive black hole shadow of  M87*  observed by the EHT. These results may provide a new path for choosing a suitable modified theory of gravity that is consistent with the recent observational data.

\begin{acknowledgments}

The authors
 appreciate the respectful referees for carefully reading the manuscript and their insightful comments which boosted
 the quality of the paper, considerably
\end{acknowledgments}

\end{document}